\documentstyle{article}
\begin{document}
\begin{center}
{\Large \bf Quantum Cable as transport spectroscopy of 1D DOS of
cylindrical quantum wires}
\end{center}

\vspace{.3cm}
\noindent
\begin{center}
{\large Y. Xiang, Z. Y. Zeng$^{*}$, and L. D. Zhang}
\end{center}

\vspace{.001cm}
\noindent
\begin{flushleft}
{\it Institute of Solid State Physics, Chinese Academy of
Sciences, \rm P.O. Box 1129, Hefei,\\ 230031, People's Republic of China \\}
\end{flushleft}

\vspace{.5cm}
\noindent
\begin{center}
{\bf Abstract}
\end{center}

We considered the proposed Quantum Cable as a kind of transport
spectroscopy of one-dimensional (1D) density of states (DOS) of
cylindrical quantum wires. By simultaneously detecting the direct
current through the cylindrical quantum wire and the leaked
tunneling current into the neighboring wire at desired temperatures,
one can obtain detailed information about 1D DOS and subband structure
of cylindrical quantum wires.

\vspace{1.cm}

\vspace{5cm}
\noindent
* Correspondence to whom should be addressed.

\noindent
Tel: 0551-5591444 (office), 5591453 (home)

\noindent
Fax: 0551-5591434

\noindent
E-mail:zyzeng@mail.issp.ac.cn

\newpage

\noindent

   Nanometer structure and materials are of great interest as potential
candidates for future electronic devices of greatly reduced
size$^1$. The fundamental physics underlying these devices
is quantum coherence and quantum confinement$^2$. Rational design and
fabrication will require a better understanding of how the properties
of these materials depends on, for example, their dimensionality and
size.  With the rapid development of nanofabrication and synthesis
technology, a variety of low-dimensional nanostructures with unusual
characteristics have been realized,
such as quantum well, quantum wire,
quantum dot as well as their composite structures$^3$.

   As a kind of quantum wire structure, nanotubes$^4$ have received extensive
 attention both theoretically and experimentally. Most recently, the group
 of Iijima$^5$ has successfully synthesized a new
 kind of coupled nanotube
 structure -
 coaxial nanocable, in which two conducting cylindrical layers are coupled
 through a very thin insulating layer.
 Motivated by this new nanostructure, we
 proposed Quantum Cable$^6$
 similar to coaxial nanocable structure, and have
 studied
 its energy subband spectrum and
  ballistic transport properties. Quantum Cable
 comprises two cylindrical quantum wires and a tunable potential barrier,
 which is sketched in Fig. 1 (a). It is achievable either by confining
 electrons in the nanocable-like potential wells, or
 by curling multilayer heterostructure into a solid
 cylinder or hollow cylinder with nanometer width.
 Since Quantum Cable is a coupled-quantum-wire structure,
it is very akin to the usual coupled dual-quantum-well waveguides$^7$ (DQWs)
in two-dimensional electron gas (2DEG).
 On the other hand, Quantum
 Cable is different from the DQWs due to its cylindrical symmetry$^6$.
  For this reason Quantum Cable is a unique
 nanostructure and expected to be used as some kind of quantum-effect
 device. In this letter, we considered Quantum Cable as a transport
 spectroscopy for detecting quasi-one-dimensional density of states (1D DOS)
 (bound states) in cylindrical quantum wires. The system will be studied
 is two GaAs cylinder wires surrounded by $Ga_{0.7}Al_{0.3}As$ for
 which the parameters are $U_B=100$ mev, $m_1^*=5.73\times 10^{-32}
 kg$ and $m_2^*=1.4m^*_1$. The widths of two cylindrical wires are
 both $50$ nm and the barrier width is set to be $5$ nm.

   The bias configuration [see Fig. 1 (b)]
   is as follows: a small constant bias
   $V_D$ ( 100 $\mu$v) drives
electrons' direct flowing  through one of the cylindrical quantum wires
into the drain, while another variable bais voltage $V$
is applied to the side-wall of
the same quantum
wire to adjust the electrons' Fermi potential.
In the forthcoming
calculations, we directly relate the applied side-wall voltage to
the Fermi potential of electrons in quantum cylinder for simplicity.
In fact, this needs self-consistent calculation. Note that such
simplication will not lead to any confusion.
The potential barrier
is set to be good isolation for qauntum wires, yet allow for electron's
tunneling between two wires. Then tunneling current through the barrier
can be measured simultaneously
in another wire. From Landauer-B{\"u}ttiker
formula$^8$,  the direct current $I_D$ along the cylindrical
wire from source to drain can be derived as
\begin{equation}
I_D=\frac{2e^2}{h} \sum_{mn}\int_{E_{mn}}^{\infty}dE
[-\frac{\partial f(E-E_F-eV)}
{\partial E}]V_D,
\end{equation}
and the tunneling current $I_T$ from one cylinder wire to another
is given by
\begin{equation}
I_T=e\sum_{mn}\int_{-\infty}^{+\infty}dE v_{\bot}
(E-E_{mn})g_{mn}^{1D}(E-E_{mn})
T_{mn}(E-U_B)[f(E-E_F-eV)-f(E-E_F)],
\end {equation}
where we have accounted for the contribution to
the current from every occupied subband $(mn)$.
$E_F$ is electrons' Fermi energy as $V=0$, and will be set to be
zero for convinience in our actual numeration, $E_{mn}$ is the energy to be
determined of the $(mn)th$
subband and $U_B$ is the height of the tunneling barrier. $f(E)=
[1+e^{E/(k_bT)}]^{-1}$ is the Fermi distribution function with
$k_b$ the Boltzmann constant.
 $v_{\bot}(E-E_{mn})=
(2E_{mn})^{1/2}/(2m_1^*)$ is the velocity of tunneling electrons at subband
(mn).
$g_{mn}^{1D}(E-E_{mn})=\theta(E-E_{mn})(E-E_{mn})^{-1/2}/(\pi \hbar)$
is the 1D DOS. The transmission
coefficient is approximated to be that through a 1D square barrier, i.e.,
$T(E-U_B)=[1+sh^2(\sqrt{2m_2^*(U_B-E)}R_B/\hbar)/(E/U_B-E^2/U_B^2)]^{-1}$
$ (E<U_B)$ .
The subband
energy $E_{mn}$ is followed from
\begin{equation}
J_n(\sqrt{2m_1^*E}R_1/\hbar)Y_n(\sqrt{2m_1^*E}R_2/\hbar)-
Y_n(\sqrt{2m_1^*E}R_1/\hbar)J_n(\sqrt{2m_1^*E}R_2/\hbar)=0,
\end {equation}
where $J_n(x), Y_n(x)$ are the first-order,second-order Bessel
functions, $R_1$, $R_2$ are the inner and outer radius of the
hollow cylindrical quantum wire considered, respectively.

    Once the subband energy is obtained, the 1D DOS of the cylindrical
wire can be calculated. In Fig. 2 we present the calculated 1D DOS
for a hollow cylindrical quantum wire of width $50$ nm ,
in the approximation of infinite potential well.
It is worth noting that subband spectrums are the same in the cases of
both infinite
and finite confining potential well model,
except that subband energy in the case of
infinite
potential well is slightly higher than that in the finite potential well
case for not very small wire width$^9$. From Fig. 2, one can find that,
over the consecutive energy range, subband structure and 1D DOS of
a hollow quantum cylinder are
very similar to that of a single 2D quantum waveguide (SQW).
Whereas irregular subband arrangement appears in the intersection
between two energy regions with SQW-like subband structure. This
feature arises from the mixing of subbands of
lowest-order azimuthal quantum number $m$ and higher-order
radial quantum number $n$ with that of large $m$ and small $n$.
For solid cylinder,
its energy subbands are organized somewhat irregularly$^6$.

In Fig. 3, we give the direct current $I_D$ through a quantum cylinder
and tunneling current $I_T$
leaked into another cylinder at absolute zero temperature,
 as the side-wall voltage
bias is adjusted. It can be shown that the direct and tunneling currents
decrease as $V$ is made smaller, until at $V < 2.2$ mv, all current-carrying
channels (subbands) are
turned off and no direct and tunneling current is observed.
If one examines carefully
the detailed structure of direct and tunneling currents, one can see
the perfect quantized steps in the direct current $I_D$,
 and the 1D DOS-similar
peak oscillation in the tunneling current $I_T$.
Moreover, the start of each step in $I_D$ lines perfectly up with
the corresponding peak in $I_T$, as indicated by the arrows in
Fig. 3. By counting the number of
steps in $I_D$ and  peaks in $I_T$, we find they are the same.
In addition,
two steps of conductance magnitude $2e^2/h$
can be discerned from the other steps of conductance unit $4e^2/h$.
The former steps come from the contribution of the conducting subbands
belonging
to the azimuthal quantum number $m=0$, while the later result from
the conducting subbands $m \neq 0$ which are doubly degenerated.
This reflects the cylindrical symmetry of cylindrical quantum
wires.
As the side-wall voltage increases, electron's Fermi
potential sweep the transverse energy subband one by one and thus
conducting channels, each of which
carrying the same ammount of current, opens up progressively.
Then we can observe quantum steps in $I_D-V$ characteristic.
As to the peak structure in the tunneling current $I_T$, it is
the direct consequence of tunneling current being propotional
to the 1D DOS for a given subband, as can be found in Eq. (2).

At nonzero temperature, it is expected that
the plateaus in the direct current will be rounded
and the peak in tunneling current will be  broadened,
until these steps and peaks are smeared out by thermal effect.
Fig. 4 shows the currents $I_D$ and $I_T$ vs side-wall bias
at temperatures $T=0.2, 1, 3.5$ K. It is evident that,
with the increase of the temperature, the plateaus in $I_D$
are more rounded and the peaks in $I_T$ more broadened.
As the thermal energy is comparative to the energy interval
between two adjoining subbands, the original two distinct steps
at low temperature will be blurred, and two narrow peaks will be mingled
into one broadened peak.
At around $T=3.5$ K, the step-fasion structure in
$I_D$ as well as the peak oscillations of $I_T$ is washed out. This is
consistent with the predicted smearing-out temperature
by Bagwell and Orlando$^{10}$. If the width of cylindrical
quantum wires is decreased, intersubband energy spacing
will increase, which implies a higher operatation temperature,
though the quantum steps in direct current still will be
smeared out at about $T=3.5$ K becuase of fixed height of current step.
As a result, it is more convinient and
more accuate to determine the 1D DOS and subband
structure of cylindrical
quantum wire from the leaked tunneling
current than from the direct current into the drain. However,
one can not gain detailed insight into the 1D DOS and subband structure
of cylindrical quantum wire simply from the data of tunneling current.
Since at some temperature, two narrow
peaks in tunneling current
corresponding to the neighboring subbands of small energy
interval will be smeared into one broadened peak, at this moment one can
notice this fact according to the slope of the direct current at
corresponding side-wall bias, and make good resolution
of such neighboring subbands .
Therefore it is necessry to measure simultaneous the
direct current in the drain and leaked current into another
cylindrical quantum wire.

Based on the aforementioned observations, one can determine the energy
subband structure and 1D DOS of cylindrical quantum wire from the
simultaneous measurement of the direct current through the wire and
the leaked current into the neighboring cylinder wire at accessible
temperatures.  To do so, one should first form
Quantum Cable structure from a single cylindrical
quantum wire using either fabrication or synthesis approach. Then apply
bias voltages with the
configuration mentioned above to the obtained Qauntum Cable.
By simultaneous measuring the direct current and tunneling current
at desired temperature, one can capture detailed information about
the subband structure and 1D DOS of a cylindrical quantum wire.
Experimentally, such transport spectroscopy is feasible, since
two-terminal$^{11}$
and multi-terminal$^{12}$ measurement of probing electriccal transport in
individual single-wall carbon nanotubes have been achieved.

  This work is supported by a key project for fundamental research
in the National Climbing Program of China, one of us (ZYZENG) would like
to thank Prof. G. H. Li, Prof. G. W. Meng, Dr. Z. Cui
 for helpful discussions.

\vspace{.5cm}
\noindent
{\bf References}

\vspace{.1cm}
\noindent
\begin{enumerate}
\item T. W. Ebbesen, Annu. Rev. Mater. Sci. {\bf 24}, 235 (1994);
C. -G. Wu and T. Bein, Science {\bf 266}, 1013 (1994).
\item B. L. Altshuler, P. A. Lee, and R. A. Webb, Eds., {\it Mesoscopic
Phenomenon in Solids} (North-Holland, Amsterdam, 1991).
\item M. A. Reed, ed., {\it Nanostructured Systems} (Academic Press,
New York, 1992).
\item S. Iijima, Nature {\bf 354}, 54 (1991);
M. S. Dresseelhaus, G. Dresseelhaus, and P. C. Eklund,
{\it Science of Fullerenes and Carbon Nanotube} (Academic, San Diego, 1996)
and references cited.
\item X. Suenaga, C. Colliex,  N. Demoncy, A. Loiseau, H. Pascard,  and
F. Willaime, Science {\bf 278}, 653 (1997);
Y. Zhang, K. Suenaga, C. Colliex and S. Iijima,   Science {\bf 281},
 973 (1998).
\item Z. Y. Zeng, Y. Xiang, and L. D. Zhang (unpublished).
\item
J. A. del Alamo and C. C. Eugster, Appl. Phys. Lett. {\bf 56}, 78 (1990);
Phys. Rev. Lett. {\bf 67}, 3586 (1991).
\item R. Landauer, Philos. Mag. {\bf 21}, 863 (1970);
M. B{\" u}ttiker, Phys. Rev. Lett. {\bf 57}, 1761 (1986).
\item N. C. Constantinou and B. K. Ridley,
J. Phys.:Condens. Matter {\bf 1},2283 (1989).
\item P. F. Bagwell and T. P. Orlando, Phys. REv. B {\bf 40},
1456 (1989).
\item H. Dai, E. W. Wong, and C. M. Lieber, Science {\bf 272},
523 (1996); S. J. Tans, M. H. Devoret, H. Dai, A. Thess, R. E. Smalley,
L. J. Geerliges, and C. Dekker, Nature {\bf 386}, 474 (1997).
\item A. Bezryadin, A. R. M. Verschueren, S. J. Tans,
and C. Dekker, Phys. Rev. Lett. {\bf 80}, 4036 (1998).
\end{enumerate}

\newpage
\noindent
\begin{center}
{\bf Figure Captions}
\end{center}
\vspace{0.2cm}
\noindent
{\bf Fig.~1}~ (a) Schematic view of Quantum Cable structure, (b)
bias configuration to the probing of direct and tunneling currents
in Quantum Cable.

\vspace{0.2cm}
\noindent
{\bf Fig.~2}~   1D DOS of hollow quantum cylinders with width $50$ nm in
the approximation of infinite potential well.

\vspace{0.2cm}
\noindent
{\bf Fig.~3}~    Variation of the direct current $I_D$
through the quantum cylinder and tunneling current $I_T$ as the side-wall
voltage bias is altered, at absolute zero temperature.

\vspace{0.2cm}
\noindent
{\bf Fig.~4}~    Variation of the direct current $I_D$
through the quantum cylinder and tunneling current $I_T$ as the side-wall
voltage bias is altered, at nonzero temperatures.

\end {document}
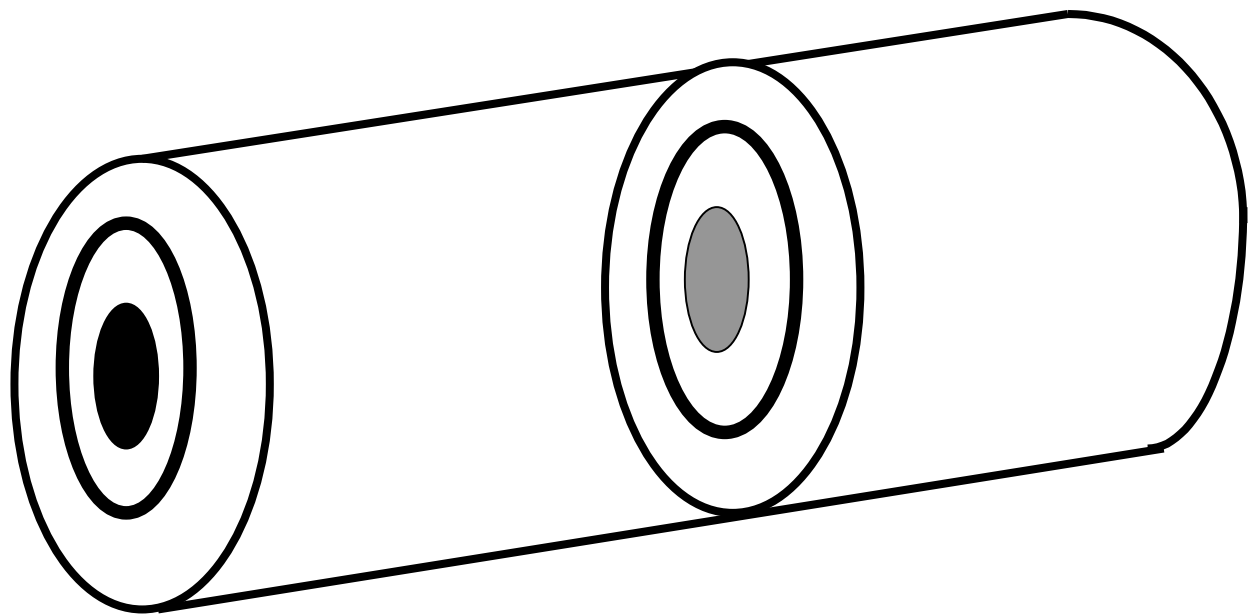
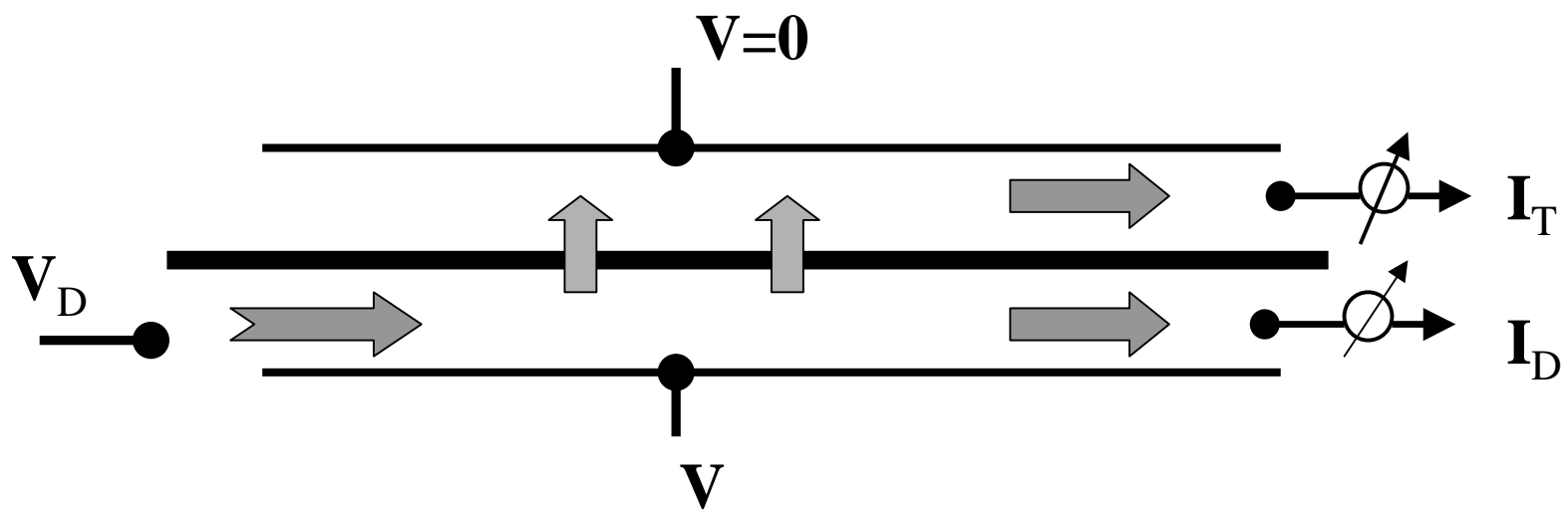
